\documentstyle[11pt,epsfig,aaspp4]{article}

\received{ }
\accepted{ }

\lefthead{Catanese et al.}
\righthead{Multiwavelength Observations of Markarian 501}

\begin{document}

\title{Multiwavelength Observations of a Flare from Markarian 501}

\author{M. Catanese,\altaffilmark{1}
S. M. Bradbury,\altaffilmark{2}
A. C. Breslin,\altaffilmark{3}
J. H. Buckley,\altaffilmark{4}
D. A. Carter-Lewis,\altaffilmark{1}
M. F. Cawley,\altaffilmark{5}
C. D. Dermer,\altaffilmark{6}
D. J. Fegan,\altaffilmark{3}
J. P. Finley,\altaffilmark{7}
J. A. Gaidos,\altaffilmark{7}
A. M. Hillas,\altaffilmark{2}
W. N. Johnson,\altaffilmark{6}
F. Krennrich,\altaffilmark{1}
R. C. Lamb,\altaffilmark{8}
R. W. Lessard,\altaffilmark{7}
D. J. Macomb,\altaffilmark{9,10}
J. E. McEnery,\altaffilmark{3}
P. Moriarty,\altaffilmark{11}
J. Quinn,\altaffilmark{3,4}
A. J. Rodgers,\altaffilmark{2}
H. J. Rose,\altaffilmark{2}
F. W. Samuelson,\altaffilmark{1}
G. H. Sembroski,\altaffilmark{7}
R. Srinivasan,\altaffilmark{7}
T. C. Weekes,\altaffilmark{4}
and J. Zweerink\altaffilmark{1}
}

\altaffiltext{1}{Department of Physics and Astronomy, Iowa State
University, Ames, IA 50011-3160}
\altaffiltext{2}{Department of Physics, University of Leeds,
Leeds, LS2 9JT, Yorkshire, England, UK}
\altaffiltext{3}{Experimental Physics Department, University College, 
Belfield, Dublin
4, Ireland}
\altaffiltext{4}{ Fred Lawrence Whipple Observatory, Harvard-Smithsonian 
CfA, P.O. Box 97, Amado, AZ 85645-0097} 
\altaffiltext{5}{Physics Department, St.Patrick's College,
Maynooth, County Kildare, Ireland}
\altaffiltext{6}{E. O. Hulburt Center for Space Research, Code 7650,
Naval Research Laboratory, Washington, D.C. 20375}
\altaffiltext{7}{Department of Physics, Purdue University, West
Lafayette, IN 47907}
\altaffiltext{8}{Space Radiation Laboratory, California Institute of
Technology, Pasadena, CA 91125}
\altaffiltext{9}{NASA/Goddard Flight Center, Code 662, Greenbelt, 
MD 20771}
\altaffiltext{10}{Universities Space Research Association}
\altaffiltext{11}{Department of Physical Sciences, Regional Technical College,
Galway, Ireland}

\authoremail{catanese@egret.sao.arizona.edu}
\authoraddr{Michael Catanese, F. L. Whipple Observatory, P.O. Box 97, 
Amado, AZ 85645}

\begin{abstract} 

We present multiwavelength observations of the BL Lacertae object
Markarian 501 (Mrk 501) in 1997 between April 8 and April 19.
Evidence of correlated variability is seen in very high energy (VHE,
$E \gtrsim$ 350 GeV) $\gamma$-ray observations taken with the Whipple
Observatory $\gamma$-ray telescope, data from the Oriented
Scintillation Spectrometer Experiment (OSSE) of the {\sl Compton
Gamma-Ray Observatory} ({\sl CGRO}), and quicklook results from the
All-Sky Monitor (ASM) of the {\sl Rossi X-ray Timing Explorer} ({\sl
RXTE}) while the Energetic Gamma-Ray Experiment Telescope (EGRET) did
not detect Mrk 501.  Short term optical correlations are not
conclusive but the U-band flux observed with the 1.2m telescope of the
Whipple Observatory was 10\% higher than in March.  The average energy
output of Mrk~501 appears to peak in the 2 keV to 100 keV range
suggesting an extension of the synchrotron emission to at least 100
keV, the highest observed in a blazar and $\sim$100 times higher than that
seen in the other TeV-emitting BL Lac object, Mrk 421.  The VHE
$\gamma$-ray flux observed during this period is the highest ever
detected from this object.  The VHE $\gamma$-ray energy output is
somewhat lower than the 2-100 keV range but the variability amplitude
is larger.  The correlations seen here do not require relativistic
beaming of the emission unless the VHE spectrum extends to $\gtrsim$5
TeV.

\end{abstract}


\section{Introduction}
\label{intro}

Markarian 501 (Mrk 501) is a BL Lacertae object (BL Lac) at $z=0.033$
(e.g., \cite{Stickel93}) making it the second closest known BL Lac
after Mrk 421.  
Mrk 501 was discovered as a $\gamma$-ray source at $E >$ 300 GeV by the
Whipple collaboration (\cite{Quinn96}) and recently confirmed by the
HEGRA collaboration (\cite{Bradbury97}).  Its initial detection level
was 8\% of the VHE flux from the Crab Nebula.
Like the other TeV-emitting
AGN, Mrk 421, Mrk 501 exhibits day-scale variability in its VHE
$\gamma$-ray emission (\cite{Quinn96}), but prior to 1997, the $\gamma$-ray
emission from Mrk 421 was generally observed to have a higher mean
flux (\cite{Schubnell96}), to be more frequently variable
(\cite{Buckley96}), and exhibit higher amplitude and shorter
time-scale flares (\cite{Gaidos96}).  EGRET has never detected Mrk 501
(\cite{Fichtel94}; \cite{Thompson95}).  Its low VHE flux and the
nearby EGRET source, 4C~38.41 (\cite{Mattox93}),
were previously assumed to account for this.

Here, we present Whipple Observatory VHE $\gamma$-ray and optical, OSSE,
and EGRET observations of Mrk 501 during a flare in 1997 April.  We
also include quicklook X-ray results from the ASM.  Overlapping
observations of Mrk 501 with {\sl RXTE} (\cite{Catanese97}) and {\sl
Beppo SAX} (\cite{Pian97}) are reported elsewhere.  Details of the
observations are given in \S~2.  They show evidence of a correlation
among the wavebands covered by the above experiments (see \S~3),
except EGRET which detected no emission from Mrk 501.  The
implications of these results are discussed in \S~4.

\section{Observations and Analysis}
\label{observe}

\subsection{Whipple Observatory observations}
\label{whipple}

The VHE observations reported in this paper were made with the
atmospheric \v{C}erenkov imaging technique (\cite{Cawley95}) using the
10~m optical reflector located at the Whipple Observatory on Mt.
Hopkins in Arizona (elevation 2.3 km) (\cite{Cawley90}).  A 151
photomultiplier tube (upgraded in 1996 from 109) camera mounted in the
focal plane of the reflector records images of atmospheric
\v{C}erenkov radiation from air showers produced by $\gamma$-rays and
cosmic rays.  The
$\gamma$-ray selection was based on the Supercuts criteria (Reynolds
et al.  1993), however the camera upgrade 
necessitated that the analysis cuts be changed slightly (\cite{Weekes97}).
The data reported in this paper are analyzed with a {\sl Tracking}
analysis (cf. \cite{Quinn96}).

\begin{figure} 
\centerline{\epsfig{file=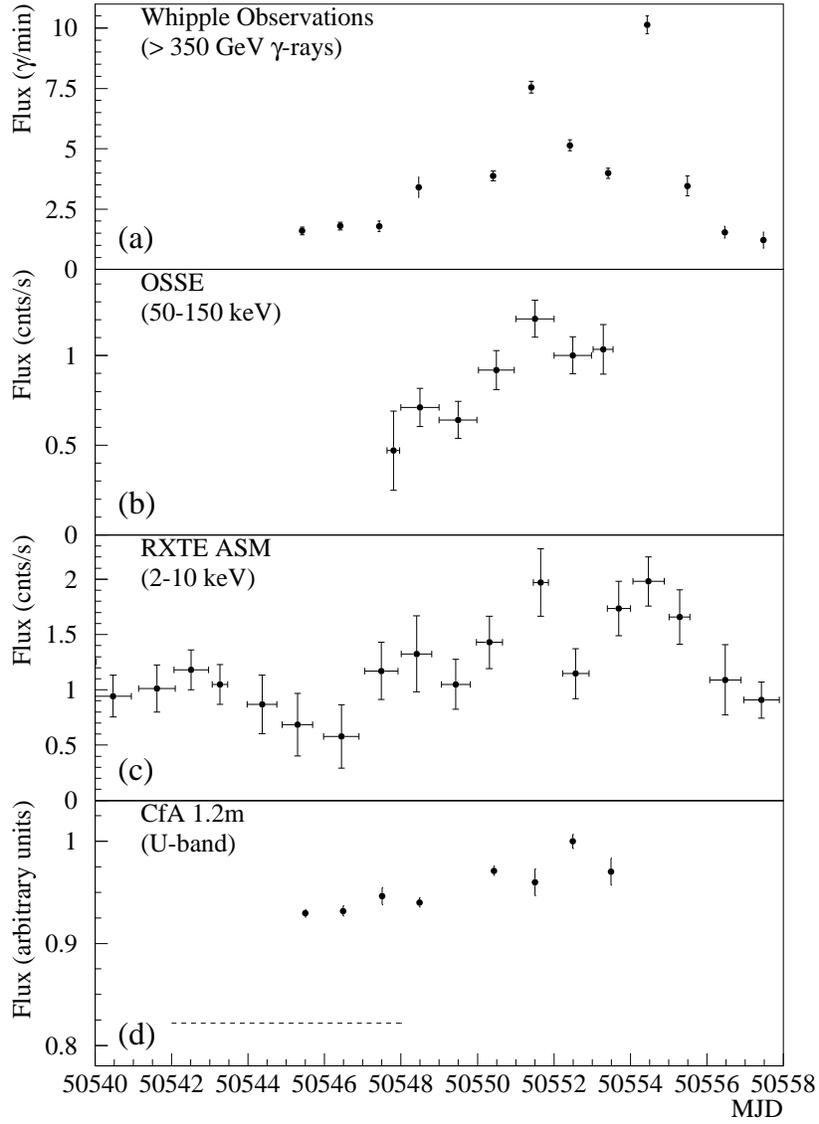,height=6in,angle=0.}}
\caption{(a) VHE $\gamma$-ray, (b) OSSE 50 keV - 150 keV, (c) ASM 2-10 keV, 
and (d) U-band optical light-curves of 
Mrk 501 for the period 1997 April 2 (MJD 50540) to 
1997 April 20 (MJD 50558).  The dashed line in (d) indicates the
average U-band flux in 1997 March.
\label{lc}
}
\end{figure}

Count rates for $\gamma$-rays are converted to integral fluxes
by expressing them as a multiple of the Crab Nebula count rate and
then multiplying that fraction by the Crab Nebula flux: 
$I(>350 \ {\rm GeV})=(1.05 \pm 0.24) \times 10^{-10}$ photons
cm$^{-2}$ s$^{-1}$ (\cite{Hillas97}).  The errors in the absolute
fluxes listed below are in general dominated by the uncertainty in the
Crab Nebula flux.  The statistical significance of the detections are
more accurately indicated by the $\gamma$-ray count rates (see 
Fig.~\ref{lc}a).  This procedure assumes that the VHE
$\gamma$-ray flux of the 
Crab Nebula is steady, as 7 years of Whipple Observatory data
indicate (\cite{Hillas97}), and that the Mrk 501 spectrum
is identical to that of the Crab Nebula between 0.3 and 10 TeV, $dN/dE
\propto E^{-2.5}$ (\cite{Hillas97}), which may not be the case.
Estimates of the VHE spectrum of Markarian 501 will be undertaken in
future work.

VHE observations of Mrk 501 were
taken nightly April 7 -- 19, for a total exposure of 19.9 hours.  
The mean flux above 350 GeV
during this period was (16.4$\pm$3.9)$\times$10$^{-11}$ photons
cm$^{-2}$ s$^{-1}$, which is $\sim$1.6 times the flux of the
Crab Nebula.  By comparison, the $\gamma$-ray flux from Mrk 501 was
never recorded to be higher than 0.5 times the Crab Nebula flux in
previous seasons (\cite{Quinn96}).  The flux during
April ranged from a low of (4.9$\pm$1.8)$\times$10$^{-11}$ photons
cm$^{-2}$ s$^{-1}$ on April 19 to a high of
(40.5$\pm$9.6)$\times$10$^{-11}$ photons cm$^{-2}$ s$^{-1}$ on 
April 16.  The latter is the highest flux recorded from Mrk 501 by the
Whipple Observatory.  No evidence of hour-scale variability was found
within this data set.

Optical observations were taken April 7 -- 15 with the 1.2m
Ritchey-Chr\'{e}tien telescope of the Whipple Observatory using
standard UBVRI filters.  The 12\arcmin \ full field CCD frames were
analyzed using relative photometry with a 6\arcsec \ aperture applied
to Mrk~501 and several comparison stars in the field of view.
The U-band fluxes (see
Fig.~\ref{lc}d) are expressed in arbitrary units with no galaxy light
subtraction.  These results are described in detail by Buckley \&
McEnery (1997).

\subsection{Compton Gamma-Ray Observatory Observations}
\label{OSSE}

{\sl CGRO} observed Mrk~501 for the period April 9 -- 15 as the
result of a Target of Opportunity initiated in response to the
reported high level of VHE $\gamma$-ray activity (\cite{Breslin97}).
We report here on the measurements of the OSSE and EGRET instruments
on {\sl CGRO}.

OSSE observes $\gamma$-rays in the 0.05 -- 10 MeV energy range.  
A description of OSSE, its performance and analysis procedures is given
by Johnson et al. (1993).  For the observation of Mrk~501, the {\sl
CGRO} viewing attitude and the OSSE observing sequence were selected
to avoid the X-ray binary Hercules X-1 and the EGRET source 4C~38.41.
This was achieved by viewing Mrk~501 $\sim 0\fdg 6$ off axis, or at
$\sim 80\%$ of maximum response, and by making background observations
which were offset by $\pm 7\fdg 5$.

The OSSE data taken during this 6 day period yielded a very strong
detection of Mrk~501 (see Table~\ref{osseflux}).  
The daily flux observed by OSSE
varied by over a factor of 2, peaking on April 13 (see Fig.~\ref{lc}).
Mrk~501 has not previously been observed by OSSE, so no comparison to
previous flux levels can be made.
The average $50 - 150$ keV flux detected from Mrk~501 is higher than
the emission seen by OSSE from any other blazar
(\cite{McNaron-Brown95}) except the highest recorded state from 
3C~273 (\cite{McNaron-Brown96}).  We note that it 
may not be strictly valid 
to compare Mrk 501 to 3C~273 because the latter object 
is not a typical blazar.
The average of the OSSE
data for this 6 day observation is well fit by a single power law in
the energy range 0.05 -- 2.0 MeV ($\chi^2/$dof$=0.794$, 64 dof) of the
form: $$\mbox{N}(E) = (1.8 \pm 0.1) \times 10^{-2} \left(\frac{E}{0.1
{\rm MeV}} \right)^{-2.08 \pm 0.15} \mbox{photons cm}^{-2} \mbox{
s}^{-1} \mbox{ MeV}^{-1}$$ where the uncertainties represent the 68\%
confidence interval for joint variation of two parameters
($\Delta\chi^2 = 2.3$).  The spectral index is somewhat steep compared
to OSSE spectra of other blazars but it is still within the
1.0 -- 2.1 range reported by McNaron-Brown et al. (1995).  There was
no statistically significant variation in the power law index with
time when fitting power law spectra to the 6 daily average OSSE
measurements.

The EGRET instrument (e.g., \cite{Kanbach88}; \cite{Thompson93}) is
sensitive to $\gamma$-rays in the energy range spanning approximately
30 MeV - 30 GeV.  The EGRET data were analyzed with maximum likelihood
techniques (\cite{Mattox96}), taking into account the contributions of
isotropic and Galactic diffuse emission (\cite{Hunter97}) and other
bright sources near the position of Mrk 501.  The observations taken
April 9 -- 15 indicate an excess of 1.5$\sigma$, which corresponds to
a 2$\sigma$ upper limit of $I$($>$100 MeV) $<$ 3.6 $\times$10$^{-7}$
photons cm$^{-2}$ s$^{-1}$.

\section{Results}
\label{results}

Figure~\ref{lc} shows daily flux levels 
for the contemporaneous observations of Mrk 501.  The average flux
level in the U-band in March is also included in the figure
(dashed line in Fig.~\ref{lc}d) to indicate the significant
($\gtrsim$10\%) increase in flux between March and April.  
An 11 day rise and
fall in flux is evident in the VHE and ASM wavebands with peaks on
April 13 and 16.  The 50-150 keV flux detected by OSSE 
also increases between April 9 and 15 with a peak on April 13.  The optical
data may show a correlated rise but the variation is small
(at most 6\%).  Subtraction of the galaxy light contribution will increase the
amplitude of this variation, but it should still remain lower than 
in X-rays given that the R-band contribution of the galaxy light
is $\sim$75\% (\cite{Wurtz96}) and the U-band contribution should be 
much less.  
The ratio of the fluxes between April 13
and April 9
are 4.2, 2.6, 1.7, and 1.01 for the VHE, OSSE, ASM, and U-band
emission respectively.

\begin{figure}
\centerline{\epsfig{file=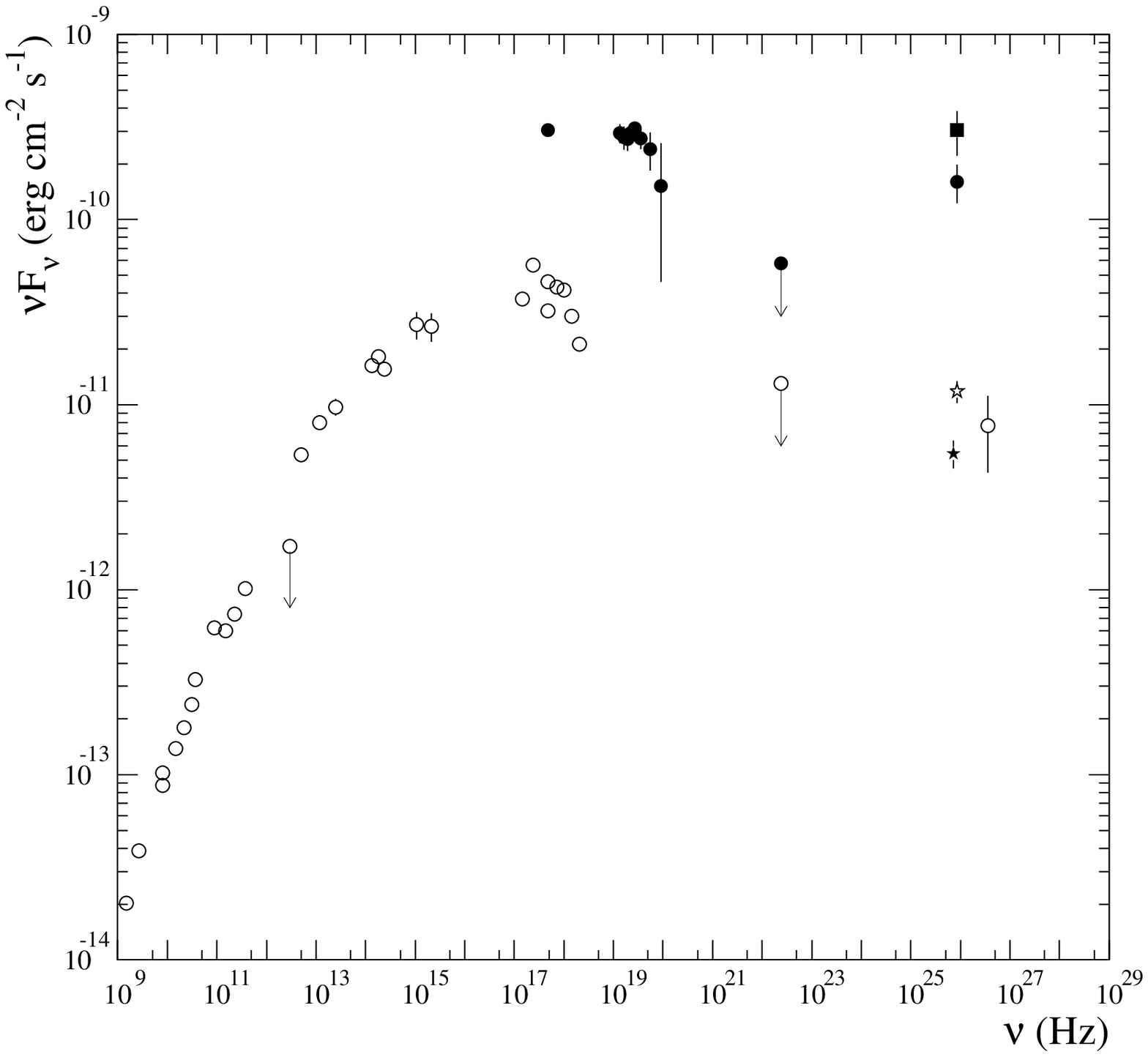,height=6in,angle=0.}}
\caption{The spectral energy distribution of Mrk 501.  The
contemporaneous observations taken April 9 -- 15 (filled circles),
archival data (open circles) (\protect\cite{Gear94};
\protect\cite{Owen80}; \protect\cite{Impey88};
\protect\cite{Kotilainen92}; \protect\cite{Edelson92};
\protect\cite{Singh85}; \protect\cite{Staubert86};
\protect\cite{Fichtel94}; \protect\cite{Bradbury97}), the mean VHE
$\gamma$-ray flux in 1995 (filled star) (\protect\cite{Quinn96}), the
mean flux in 1996 (open star), and the maximum VHE $\gamma$-ray flux,
detected on April 16, (filled square) are indicated.
\label{nufnu_plot} 
}
\end{figure}

In Figure~\ref{nufnu_plot} we show the spectral energy distribution,
plotted as the power per logarithmic bandwidth, $\nu F_\nu$, versus
frequency, $\nu$.  The average flux for April 9 -- 
15, the duration of the {\sl CGRO} ToO, is indicated by the
filled circles.  The filled square indicates the peak VHE $\gamma$-ray
flux from April 16.  
The OSSE data for this plot have been divided
into 8 energy bands between 50 keV and 470 keV.
No excess
emission is seen above 470 keV so those data are not shown.  
The VHE integral fluxes are converted to single point fluxes
by assuming the spectrum follows a power law with the same spectral
index as the Crab Nebula.  
The EGRET integral flux upper limit is converted
to a single point flux limit assuming $dN/dE
\propto E^{-2}$.  The ASM 2-10 keV count rate is converted to a flux
point by normalizing to the Crab flux and assuming Mrk 501 has the same
photon spectrum as the Crab ($E^{-2.05}$) within that energy range.
From Fig.~\ref{nufnu_plot}, a peak in the power output occurs within
the 2-100 keV range and the VHE $\gamma$-ray power output is much higher
than at EGRET energies but somewhat below that in the 2-100 keV range.

\section{Discussion}
\label{discuss}

The results presented here show that for Mrk 501, like Mrk 421 
(\cite{Buckley96}; \cite{Macomb95}), the
VHE $\gamma$-rays and the soft X-rays vary together, the energy budget at
X-rays and $\gamma$-rays is comparable, there is evidence of correlated
optical variability, and the cut-off in the synchrotron emission occurs
at much higher energies than in radio-selected BL Lacs or
flat-spectrum radio quasars.
However, Mrk~421 has a large deficit in energy
output between 1 keV and 50 keV (\cite{McNaron-Brown95};\cite{Buckley96}).  
With Mrk 501,
there is instead a peak in the power output at $\sim$100 keV, and the
deficit has shifted to between 1 MeV and 350 GeV.  
Also, the variability amplitude for the VHE
$\gamma$-rays is larger than the X-ray and OSSE variations, whereas
similar flares in Mrk 421 showed comparable amplitude variations
(\cite{Buckley96}; \cite{Macomb95}).

A likely
explanation of the OSSE detection is that the synchrotron emission in
Mrk 501 extends to 100 keV, compared with the $\sim$1 keV cutoff seen
in Mrk 421.  If this is 
synchrotron emission from nonthermal electrons, then the maximum
electron Lorentz factor is given by $\gamma_{\rm max} \cong 3\times
10^6 [ E_{\rm syn}(100~{\rm keV})]/B\delta]^{1/2}$, where $B$ is the
magnetic field strength in Gauss, $\delta$ is the Doppler factor of
the radiating plasma, and $E_{\rm syn}(100~{\rm keV})$ is the cutoff
energy of the synchrotron spectrum in units of 100 keV.  The
correlated variability between the hard X-rays and the VHE
$\gamma$-rays implies that the same nonthermal electrons are
responsible for the two emission components (cf. \cite{Macomb95}),
with the VHE $\gamma$-rays produced by Compton scattering of either
internal synchrotron photons (e.g., \cite{Konigl81}; Maraschi,
Ghisellini, \& Celotti 1992; \cite{Bloom96}) or photons produced
external to the jet (e.g., Dermer, Schickeiser, \& Mastichiadis 1992;
\cite{Blandford93}; Sikora, Begelman, \& Rees 1994). For low redshift
objects, electrons must have Lorentz factors $\gamma > 2\times 10^6
E_{\rm C}({\rm TeV})/\delta$ to Compton-scatter photons to energies of
$E_{\rm C}({\rm TeV})$ TeV.  Thus the magnetic field $B\lesssim 2
E_{\rm syn}(100~{\rm keV})\delta/[E_{\rm C}({\rm TeV})]^2$ 
(see also \cite{Buckley97b}).

A lower limit to the magnetic field is implied by the variability of
the synchrotron emission. Electrons cool through synchrotron losses in
the co-moving frame on a time scale $\Delta t_{\rm syn} = 7.7\times
10^8/(B^2\gamma)$ s.  The source might
intrinsically vary on a longer time scale, so 
a lower limit to the observed variability
time-scale for low-redshift objects is given by $\Delta t_{\rm obs}
\gtrsim \Delta t_{\rm syn}/\delta$.
This implies that $B \gtrsim [9000/(\gamma\delta\Delta t_{\rm
obs}({\rm days})]^{1/2}$.  This expression can be written in terms of
the observed photon energy $E_{\rm obs}(\rm eV)$ using the relation
$\nu_{\rm obs} = 2.8\times 10^6 B\gamma^2\delta$ from synchrotron
theory.  Putting the above constraints together, 
the allowed magnetic field strength of the emission
region is given by

\begin{equation}
[E_{\rm obs}({\rm eV})\delta \Delta t^2_{\rm
obs}({\rm days}) ]^{-1/3} \lesssim B \lesssim 2 E_{\rm
syn}(100~{\rm keV})\delta/[E_{\rm C}({\rm TeV})]^2\;.
\label{blim}
\end{equation}

\noindent The ASM ($E_{\rm obs} = 2000$ eV) flux varies on
time-scales at least as short as one day, implying that $B \gtrsim 0.08
\delta^{-1/3}$~G. Such a range of field strengths are expected at the
base of a jet from independent considerations (e.g., \cite{Marscher85}).  

Equation~\ref{blim} provides a new test for beaming in $\gamma$-ray
blazars, although it assumes 
that the high-energy component arises
from Compton-scattering of photons by the same nonthermal electrons
which produce the synchrotron component.  If the lower
limit in equation~\ref{blim} exceeds the upper limit when
$\delta$ is set equal to 1, then beaming is implied.  Here it is
especially crucial to determine the cutoff energy at TeV energies.  If
the high-energy spectrum of Mrk 501 continues without break to
$\gtrsim 5$ TeV, then beaming is implied.

This beaming test complements $\gamma\gamma$ transparency arguments
which also provide lower limits to $\delta$.  The $> 350$ GeV
$\gamma$-rays interact with the X-ray photons in the extreme
relativistic regime of pair production (i.e., $E_{\rm VHE} E_{\rm X}
\gg m_e^2c^4$), which has been treated by Dermer \& Gehrels (1995).
One finds that $\tau_{\gamma\gamma} \lesssim 0.1$ for this
interaction, so no beaming is required.  Improved limits to the
Doppler factor through $\gamma\gamma$ transparency arguments and from
equation~\ref{blim} will be obtained if short time scale correlations
are detected between VHE $\gamma$-rays and photons in the EUVE or
optical regime.  Also, if the spectrum observed with OSSE extends
beyond 0.511 MeV, $\gamma\gamma$ transparency arguments can be applied
to this $\gamma$-ray emission directly (\cite{McNaron-Brown95}).

Finally, it will be important to investigate whether the extension of
the synchrotron emission in Mrk 501 to 100 keV is always present, or
only occurred during the flaring activity shown in this work.
Previous compilations of X-ray spectra for Mrk 501 spanning 0.03 - 30
keV (e.g., \cite{Ciliegi93}) are not conclusive, mainly due to the low
statistics at the high energies.  Further studies of Mrk 501 with OSSE
and the high energy instruments on {\sl RXTE} and {\sl BeppoSAX} could
resolve this issue.

\acknowledgments 

We acknowledge the technical assistance of K. Harris.  This research
is supported by grants from the U. S. Department of Energy and by
NASA, by PPARC in the UK and by Forbairt in Ireland.  The X-ray data
in this work are quicklook results presented by the ASM/RXTE team.

\begin{deluxetable}{cccc}
\tablewidth{0pt}
\tablecaption{OSSE flux measurements for Mrk 501 \label{osseflux}}
\tablehead{\colhead{ } & \colhead{0.05-0.15 MeV} & 
\colhead{0.15-0.50 MeV} & \colhead{0.5-3.5 MeV}} 
\startdata
Flux\tablenotemark{a} & 24.2$\pm$1.3 & 2.0$\pm$0.3 & $<$0.2 \nl
\enddata
\tablenotetext{a}{Flux units are 10$^{-3}$ photons cm$^{-2}$ s$^{-1}$ 
MeV$^{-1}$.  The upper limit is 2$\sigma$.}
\end{deluxetable}

\end{document}